\documentclass[astrosymb]{aastex631}

\newcommand{\gtsimeq}{\raisebox{-0.6ex}{$\,\stackrel
    {\raisebox{-.2ex}{$\textstyle >$}}{\sim}\,$}}

\newcommand{\Mearth}{\mbox{$M_{\oplus}$}}
\newcommand{\change}[1]{{#1}}
\newcommand{\ms}{\mbox{m\,s$^{-1}$}}

\submitjournal{AJ}
\accepted{October 31, 2024}

\graphicspath{{./}{}}

\begin{document}

\title{The Search for the Inbetweeners: How packed are \textit{TESS} planetary systems?}

\correspondingauthor{Jonti Horner}
\email{jonti.horner@unisq.edu.au}

\author[0000-0002-1160-7970]{Jonathan Horner}
\affiliation{Centre for Astrophysics, University of Southern Queensland, West Street, Toowoomba, QLD 4350, Australia}

\author[0000-0001-9957-9304]{Robert A. Wittenmyer}
\affiliation{Centre for Astrophysics, University of Southern Queensland, West Street, Toowoomba, QLD 4350, Australia}

\author[0000-0002-7084-0529]{Stephen R. Kane}
\affiliation{Department of Earth and Planetary Sciences, University of California, Riverside, CA 92521, USA}
\affiliation{Centre for Astrophysics, University of Southern Queensland, West Street, Toowoomba, QLD 4350, Australia}

\author[0000-0003-0437-3296]{Timothy R. Holt}
\affiliation{Centre for Astrophysics, University of Southern Queensland, West Street, Toowoomba, QLD 4350, Australia}
\begin{abstract}

In this work, we examine seven systems discovered by \textit{TESS}, to see whether there is any room in those systems for an additional planet (or several) to lurk unseen between the two planets already confirmed therein. In five of those systems (namely HD~15337; HD~21749; HD~63433; HD~73583 and LTT~3780) we find that there is ample room for an undiscovered planet to move between those that have already been discovered. In other words, as they currently stand, those systems are \textbf{not tightly packed}. In stark contrast, the perturbative influence of the two known TOI-1670 planets is such that additional planets in between are ruled out. The final system, TOI~421, is more challenging. In the vast majority of cases, adding an Earth-mass planet to that system between the orbits of the known planets caused catastrophic instability. Just $\sim$1.1\% of our simulations of the modified system proved dynamically stable on a timescale of one million years. As a result, it seems that there is very little room between the two known planets in the TOI~421 system for an addition unseen world to exist, but the existence of such a planet can not be definitely ruled out on dynamical grounds alone.

\end{abstract}


\section{Introduction} \label{sec:intro}
Until the final decade of the 20th Century, humanity knew just a single planetary system - the Solar system. Our entire understanding of planetary formation - and our expectations of what planetary systems around other stars would look like - was based on our knowledge of the eight planets and assorted debris orbiting the Sun. As a result, we expected the planetary systems we hoped to find orbiting other stars to resemble our own - small, rocky planets orbiting with periods of months to a couple of years, and giant behemoths, in the icy depths, with orbits measured in decades\footnote{For a detailed overview of the Solar system, and discussions of how our understanding of the nature and formation of planets has changed through the first three decades of the Exoplanet Era, we direct the interested reader to \cite{HornerSSRev}, and references therein}.

This all changed with the discovery of the first planets orbiting other stars. The first planet found around a Sun-like star - 51 Pegasi~b \citep{Mayor1995} - was dramatically different to our expectations. A Jupiter-mass planet, it moved on an orbit incredibly close to its host star, with an orbital period of just a few days, rather than a decade or more. It was clear that the planet formation models of the day \citep[as described in review by][]{Lissauer93} needed significant revision to fit with the new data - a process which continues to the current day.

At the time of writing, 5,741 exoplanets have been confirmed orbiting in 4,270 systems\footnote{Data from the NASA Exoplanet Archive, at \url{https://exoplanetarchive.ipac.caltech.edu/}, accessed on 31st July 2024.}. The vast majority of these planets were discovered by two exceptional space observatories -- \textit{Kepler} \citep[with a tally of 3,321 confirmed planets as of July 31st, 2024, e.g.][]{Bor1,Bor2,Bat13,How14} and \textit{TESS} \citep[543 confirmed planets as of July 31st, 2024, e.g.][]{Rick15,Quinn19,Addison21,Guer21}. \change{These great missions were quick to reveal giant planets moving on orbits close to their host stars due to the observational bias of the transit method \citep{kane2008b}, though only $\sim$1\% of stars host hot Jupiters (like 51 Pegasi~b) \citep[e.g.][]{HJFreq1,HJFreq2,HJFreq3}. Typically, those hot Jupiters seem to be relatively solitary beasts - even when other planets are found in those systems, they are usually well separated from the hot Jupiter \citep[e.g.][]{lonely1,lonely2,lonely3,lonely4}.}

Whilst hot Jupiters formed the vanguard of discoveries in the Exoplanet Era, as our technology has improved, we have found an ever increasing population of smaller planets. Once again, the bulk of those planets have been revealed by \textit{Kepler} and \textit{TESS}, with `super-Earths' and `mini-Neptunes' now making up the bulk of known exoplanets \citep[e.g.][]{occ1,occ2,occ3}.  A common theme of systems in which super-Earths and mini-Neptunes are found is that, where there is one planet, there are almost certainly more. Planets in this mass range tend to come with companions - which typically all move on circular (or near-circular) orbits, suggesting that those systems have not been strongly dynamically stirred \citep[such as by the inward migration of a hot Jupiter, e.g.][]{stir1,stir2,stir3}. There is a growing consensus that such systems are often `dynamically packed' - with the planets therein so close together that there is simply no room for others to be squeezed in \citep[e.g.][]{pack1,pack2,pack3}. 

In that light, it is interesting to consider systems where \textit{TESS} has identified two transiting planets. Are those systems dynamically packed, or is there room for as-yet undetected planets to lurk in the gap between the known worlds? Given the growing evidence that most \textit{TESS} systems are dynamically packed, identifying systems with gaps large enough to fit additional planets can help direct the future search for new planets to promising targets - and also help to identify the most likely orbital periods for those planets, to help focus the search for those worlds. 

\change{Several studies have, in recent years, recognised the importance of such efforts, and have attempted to identify systems in which additional undiscovered planets might be present. \cite{Ag18,Ag19} were particularly interested in the possibility of identifying systems where potentially habitable Earth-mass planets might exist in the Habitable Zone of systems already known to contain at least one massive planet, and performed dynamical simulations to test the stability of putative Earth-mass objects in such systems - identifying a number of systems in which such planets might lurk undetected, and flagging several as potentially interesting targets for future radial velocity surveys. \cite{Dynamite} employed a markedly different approach, with the introduction of the DYNAMITE package that intends to identify systems that could host additional planets between the orbits of known planets, and to predict the likely orbit, mass, and radius of those planets. Where the work of \cite{Ag18,Ag19} used direct n-body simulations, DYNAMITE employs a simple dynamical stability criterion - requiring that planets be separated by at least eight mutual Hill radii in order to remain dynamically stable. In this manner, they drastically reduce the computational overhead required to test systems for the potential to host additional planets, allowing them to far rapidly assess the potential for a given system to host an unseen planet.}

\change{In this work, we examine seven systems, discovered by \textit{TESS} over the past five years, to see whether any could host additional planets between the orbits of those already discovered}. For each system, we perform a detailed dynamical analysis, searching for regions of orbital stability between the two known planets that could host an Earth-mass object. In Section~\ref{sec:systems}, we present the seven systems considered in this work, before detailing our methodology in Section~\ref{sec:methods}. We present the results of our simulations in Section~\ref{sec:results}, before drawing our conclusions and suggesting a direction for future work in Section~\ref{sec:conclusion}.

\section{System Properties for our Sample} \label{sec:systems}
In this work, we study seven exoplanetary systems where \textit{TESS} has discovered exactly two planets on relatively short period orbits. The systems in question are HD~15337 \citep[][]{Gand19}; HD~21749 \citep[][]{Drag19}; HD~63433 \citep[][]{Mann20}; HD~73583 \citep[][]{Barr22}; LTT~3780 \citep[][]{Clout20}; TOI~1670 \citep[][]{Tran22}; and TOI~421 \citep[][]{Carl20}. Details of the masses, radii, and orbital elements of the two known planets in each of these systems are presented in Table~\ref{table:known_planets}, and the details of the planet host stars are presented in Table~\ref{table:host_stars}. 

\begin{figure*}
\centering
\includegraphics[width=1.0\textwidth]{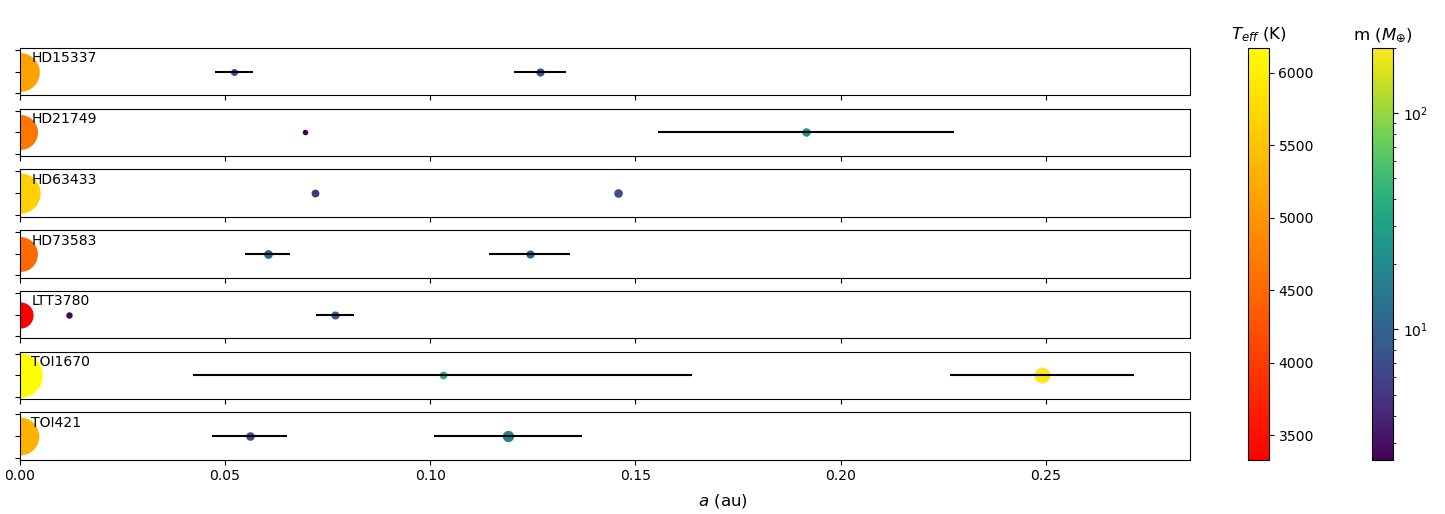} 
\includegraphics[width=1.0\textwidth]{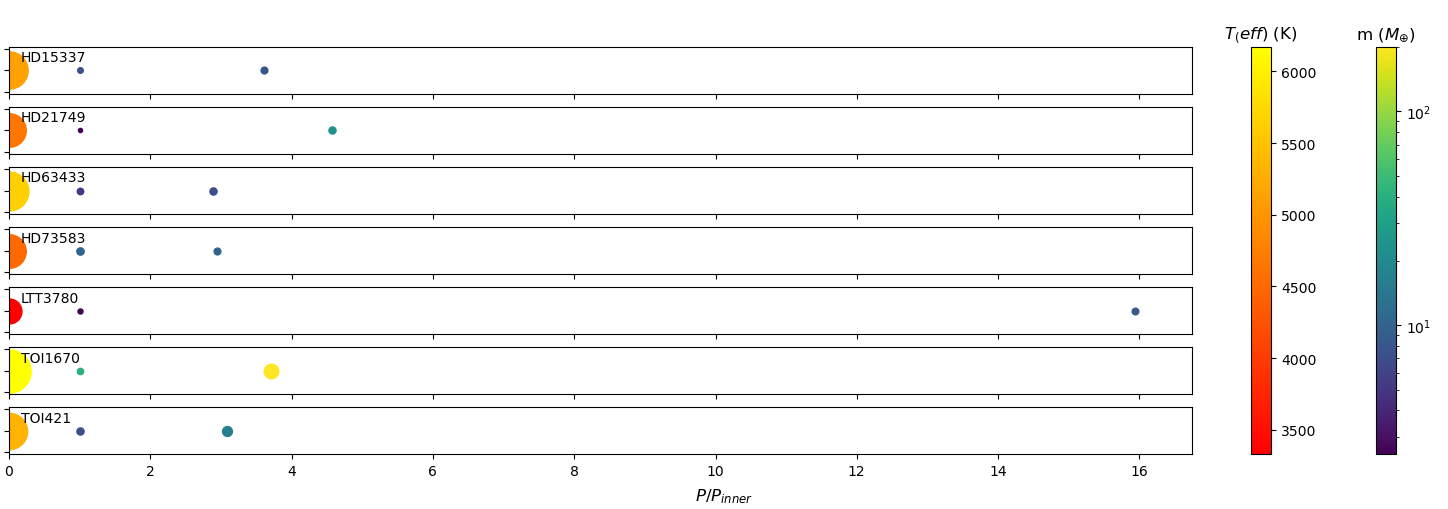} 
\caption{Schematic view of the seven planetary systems studied in this work. In the upper panel, the location on the x-axis shows the orbital semi-major axis of the planets in question, on a linear scale, with the horizontal lines reaching out from the planet showing the range of its orbit from pericentre to apocentre. The size of the marker for each planet gives the observed radius, as presented in the discovery work, whilst the colour of the marker gives the measured mass for the planet. In the case of the planets orbiting HD~63433, no measured mass has been obtained to date, and we therefore plot an estimated mass for the two planets in that system, based on Figure 3 of \cite{ChenKipping}. In the case of TOI~1670~b, the mass given is an upper limit. The lower panel presents most of the same information (mass, radius, etc.), but shows the separation of the planetary systems in terms of their orbital periods, with the period of the inner planet in each system set to unity. The parent stars are shown sized relative to one another, but not to scale with their planets, and their colour in the plot denotes their effective temperatures, $T_{\text{eff}}$.}
\label{TimPlanets}
\end{figure*}

For most of the systems studied in this work, both the mass and radii of the discovered planets are known, thanks to a combination of radial velocity (RV) and transit observations. In the case of HD~63433, however, only the radii of the planets are known -- the mass has yet to be determined. Similarly, the mass of TOI~1670~b remains unknown. \cite{Tran22} obtained RV observations of TOI~1670, but were only able to place an upper limit on the mass of TOI~1670~b of $\la 0.13 $M$_{J}$. A schematic view of the seven planetary systems is presented in Figure~\ref{TimPlanets}.
\\
\subsection{The HD~15337 system} 
HD~15337 (TOI-402) is a K1 dwarf a little older than the Sun ($\sim$5.1 Gyr old). \cite{Gand19} presented the discovery of two planets orbiting the star, with periods of ${\sim}4.76$ and ${\sim}17.19$ days, on slightly eccentric orbits. Those two planets (HD~15337~b and c) are similar in mass (${\sim}7.51$ and ${\sim}8.11$~M$_\oplus$), but markedly different in size (${\sim}1.64$ and ${\sim}2.39$~r$_\oplus$), suggesting markedly different compositions and potentially formation histories.  \change{We note that whilst this work was under review, \citet{rosario24} presented refined planetary parameters for this system which are broadly consistent with, but more precise than, those we have used here from \citet{Gand19}.}

The larger radius of HD~15337~c suggests that it is likely markedly more volatile rich than its inner companion - which in turn suggests  a possible origin at a greater orbital radius, where volatiles are more common during planet formation \citep{raymond2012,unterborn2018a}. Alternatively, the planet may have been the recipient of a significant amount of volatile material from beyond the system's snow line - which might, in turn, suggest the presence of at least one massive distant planet to dynamically source that volatile material \citep[e.g.][]{FoF1,FoF2,FoF4,FoF3,obrien2014a,raymond2017b,KW24}. A further scenario is that HD~15337~b was once markedly more volatile rich, but has been the victim of photo-evaporation \citep[e.g.][]{lopez2013,owen2013a,fulton2017}, or even that the difference between the two planets reflects a fundamentally different accretion history \citep[such as formation within an inhomogenous or poorly mixed disk; e.g.][]{zeng2016a,unterborn2019,unterborn2022}.
\\
\subsection{The HD~21749 system}
HD~21749 (TOI-186) is a K4.5 dwarf that is likely a little younger than the Sun, though its age remains extremely uncertain \citep[given as 3.8$\pm$3.7 Gyr in][]{Drag19}. The discovery of two planets orbiting HD~21749 was discovered based on \textit{TESS} observations in \cite{Drag19}. The two planets identified in that work are both significantly more massive than the Earth -- HD~21749~b has a mass of ${\sim}22.7$ M$_\oplus$, whilst HD~21749~c has a mass of ${\sim}2.5$ M$_\oplus$. Both planets have densities that suggest they are rocky or metal-rich objects ($\rho \sim$7.0 g cm$^{-3}$ for HD~21749~b and $\rho < 31.93$ g cm$^{-3}$ for HD21749~c). They move on orbits with periods of ${\sim}35.6$ and ${\sim}7.8$ days, respectively.

\cite{Drag19} note that the high density for HD~21749~b is the second highest of all planets of mass greater than 15 M$_\oplus$; the high density for HD~21749~c is the result of a mass estimated by the authors using the mass-radius relations laid out in \cite{Ning18} -- and it remains to be seen whether such a high mass and density are borne out by future observations of the system.  \change{It is worth noting that the M-R relations laid out in \citet{otegi20} and \citet{parv24} would estimate a mass of 0.8-1.5\Mearth\ for the published radius for HD\,21749~c, with a corresponding and far less extreme planetary density of 3-6 g cm$^{-3}$.}
\\
\subsection{The HD~63433 system}
HD~63433 (TOI-1726) is a young ($\sim$0.414 Gyr old) Sun-like star, of spectral class G2V. The two planets orbiting HD~63433 were announced by \cite{Mann20}, moving on orbits of period ${\sim}7.11$ and $\sim$20.5 days. Both planets are super-Earths, with radii of ${\sim}2.15$ and 2.67 r$_\oplus$. No masses were determined for the planets in that discovery work, although the authors did perform $n$-body simulations using Mercury \citep{Chambers99} to investigate the possibility of stable orbital solutions between the orbits of the two known planets. They identified an island of stability between the orbits of the two planets between 0.099 and 0.112 au that could potentially host an additional planet, suggesting that the system is a good candidate for further follow-up work in the future.  \change{While this work was under review, a third planet (HD~63433~d) was identified interior to the two considered herein \citep{cap24}. \citet{dai24} then showed that the newly discovered planet orbits in 5:3 mean-motion resonance with HD~63433~b.  That newly discovered planet is sufficiently small and distant from the gap we consider here that its presence is unlikely to influence our results.}
\\
\subsection{The HD~73583 system}
HD~73583 (TOI-560) is a young (${\sim}0.48$ Gyr) K4 dwarf. Its two known planets, HD~73583~b and c, were discovered by \cite{Barr22}, moving on orbits with period ${\sim}6.40$ and ${\sim}18.9$ days, with the discovery quickly confirmed and validated in \cite{EM23}. They are both sub-Neptunes, with radii of 2.79 and 2.39 r$_\oplus$, and masses of ${\sim}10.2$ and ${\sim}9.7$ M$_\oplus$, respectively, resulting in bulk densities of ${\sim}2.58$ and ${\sim}3.88$ g cm$^{-3}$. Such densities are sufficiently low to suggest that both planets may possess significant atmospheres, leading the authors to suggest that, since the planets are still young, they \textit{`could still [be] evolving and experiencing atmospheric mass loss'}.  Equally, the young age of the system suggests that significant quantities of volatile material might still be being delivered to these inner planets, should massive outer companions exist -- a scenario strikingly similar to models of the exogenous delivery of volatile material to the inner Solar system \citep[as discussed in][and references therein]{HJ10}.  \change{\citet{EM23} found no convincing evidence for additional planets in the system, but noted that the significant stellar activity signal has a similar periodicity to that of any planets orbiting in the gap between the two confirmed planets.} 
\\
\subsection{The LTT~3780 system}
LTT~3780 (TOI-732) is an M4 dwarf -- the least massive star of the systems studied in this work. Its age is relatively poorly constrained, having been recently estimated by \cite{Bonf24} as 3.10$_{-0.98}^{+6.20}$ Gyr. \cite{Clout20} announced the discovery of two planets orbiting LTT~3780, moving on orbits with period $\sim$0.768 and $\sim$12.3 days -- making them the most widely separated, in terms of period-ratio, of the planets studied in this work. The inner planet, LTT~3780~b, is a super-Earth, with radius ${\sim}1.33$ r$_\oplus$ and mass ${\sim}2.62$ M$_\oplus$, and a calculated bulk density of $\sim$6.1 g cm$^{-3}$ - which the authors note is \textit{`consistent with an Earth-like bulk composition'}. The outer planet, LTT~3780~c, is a  mini-Neptune, with radius $\sim$2.30 r$_\oplus$ and mass ${\sim}8.6$ M$_\oplus$.  \change{ \citet{Clout20} found no evidence for a third planetary signal in their RVs between the confirmed planets, to a limit of $K\gtsimeq$2.4\ms, or about 5\Mearth. }
\\
\subsection{The TOI~1670 system}\label{subsec:1670}
TOI~1670 is an F7 dwarf -- the most massive star of the systems examined in this work. It is $\sim$2.53 Gyr old, and hosts a warm Saturn/Jupiter (TOI~1670~c, with a mass of ${\sim}0.63$ M$_{J}$ and radius of ${\sim}0.99$ r$_{J}$). The inner of the two planets, TOI~1670~b, is most likely a super-Earth (with a measured radius of $\sim$2.06 r$_\oplus$). However, \cite{Tran22} were unable to detect TOI~1670~b in their RV dataset, meaning that they were only able to place an upper limit on the planet's mass (at $<$ 0.13 M$_{J}$). As such, it is as yet unclear whether the planet is a mini-Neptune or super-Earth. The planets move on orbits with periods of ${\sim}11.0$ days (b) and ${\sim}40.7$ days (c), corresponding to semi-major axes of ${\sim}0.103$ and ${\sim}0.249$ au, respectively. 

In addition to the masses of the planets in the TOI~1670 system being significantly larger than those of the systems considered in this work, we note that the orbit of TOI~1670~b is by far the most eccentric of any planet considered herein -- with a best-fit value of e${\sim}0.59$. \change{This likely indicates that a significant amount of dynamical evolution has occurred in the system's past, stirring the orbit of this planet \citep[e.g.][]{fordrasio08, chatterjee08, kane2014, carrera19}}. Such a high eccentricity, coupled with the high mass of the planet, likely limits the possibilities of additional undetected planets between the orbits of TOI~1670~b and c, despite their relatively wide dynamical separation.
\\
\subsection{The TOI~421 system}
TOI~421 is an ancient G9 dwarf, with an estimated age of 9.4$_{-3.1}^{+2.4}$ Gyr. It hosts two known planets -- a \textit{`super-puffy'} mini-Neptune (TOI~421~b, with a mass of $\sim$7.17 M$_\oplus$, a radius of $\sim$2.68 r$_\oplus$, and a bulk density of $\sim$2.05 g cm$^{-3}$), and a warm Neptune (TOI~421~c, mass $\sim$16.4 M$_\oplus$, radius $\sim$5.09 r$_\oplus$, and bulk density of $\sim$0.685 g cm$^{-3}$). Both planets move on moderately eccentric orbits (e~$\sim$0.163 and 0.152, respectively), with orbital periods of $\sim$5.20 and 16.1 days \citep{carleo20}. The moderate eccentricity of both planetary orbits might well indicate that a significant amount of dynamical evolution has occurred in the system in the past, and one might expect that this would impact the possibility of additional planets being found between the orbits of the two known planets in the system. \change{\citet{krenn24} presented refined orbital parameters for this system using additional \textit{TESS} and \textit{CHEOPS} transit data, and found no significant transit timing variations indicative of additional planets.}

\begin{table}[bht]
\renewcommand{\arraystretch}{1.5}
\centering
\caption{Key orbital and physical parameters for the known planets in the systems considered in this work. \label{table:known_planets}}
\begin{tabular}{cccccccc}
\hline
Planet & $P$ (d) & $a$ (au) & $e$ & $\omega$ (\degr) & $m$ ($M_{\oplus}$) & $r$ ($r_{\oplus}$) & References \\ 
\hline\hline
HD~15337~b & 4.75615$\pm$0.00017 & 0.0522$\pm0.0012$ & 0.09$\pm0.05$ & 62$_{-32}^{+42}$ & $7.51_{-1.01}^{+1.09}$ & $1.64\pm0.06$ &\cite{Gand19} \\
HD~15337~c & 17.1784$\pm$0.0016 & 0.1268$\pm0.0038$ & 0.05$_{-0.04}^{+0.06}$ & 329$_{-64}^{+69}$ & $8.11_{-1.69}^{+1.82}$ & $2.39\pm0.12$ & \\
\hline
HD~21749~b & 35.61253$_{-0.00062}^{+0.00060}$ & 0.1915$_{-0.0063}^{+0.0058}$ & 0.188$_{-0.0078}^{+0.0076}$ & 98.0$_{-17}^{+21}$ & $22.7_{-1.9}^{+2.2}$ & $2.61_{-0.16}^{+0.17}$ & \cite{Drag19} \\
HD~21749~c & 7.78993$_{-0.00044}^{+0.00051}$ & 0.0695$_{-0.0023}^{+0.0021}$ & - & - & $\sim$2.5\footnote{Mass estimated by \cite{Drag19} using mass-radius relations detailed in \cite{Ning18}; this yields an extremely high density for this planet ($\rho < 31.93$ g cm$^{-3}$), and so it seems plausible that the true mass of the planet will be markedly lower than this upper bound.} & $0.892_{-0.058}^{+0.064}$ & \\
\hline
HD~63433~b & 7.10793$_{-0.00034}^{+0.0004}$ & 0.0719$_{-0.0044}^{+0.0031}$ & 0 & 0 & - & 2.15$\pm$0.10 & \cite{Mann20} \\
HD~63433~c & 20.5453$_{-0.0013}^{+0.0012}$ & 0.1458$_{-0.0101}^{+0.0062}$ & 0 & 0 & - & 2.67$\pm$0.12 & \\
\hline
HD~73583~b & 6.3980420$_{-0.0000062}^{+0.0000067}$ & 0.0604$_{-0.0026}^{+0.0027}$ & 0.09$_{-0.06}^{+0.09}$ & 284$_{-86}^{+234}$ & 10.2$_{-3.1}^{+3.4}$ & 2.79$\pm$0.10 & \cite{Barr22} \\
HD~73583~c & 18.87974$_{-0.00074}^{+0.00086}$ & 0.1242$_{-0.0054}^{+0.0055}$ & 0.08$_{-0.06}^{+0.11}$ & 318.4$_{-47.7}^{+52.8}$ & 9.7$_{-1.7}^{+1.8}$ & 2.39$_{-0.09}^{+0.10}$ & \\
\hline
LTT~3780~b & 0.768448$_{-0.000053}^{+0.000055}$ & 0.01211$\pm$0.00012 & 0 & 251 & 2.62$_{-0.46}^{+0.48}$ & 1.332$_{-0.075}^{+0.072}$ & \cite{Clout20} \\
LTT~3780~c & 12.2519$_{-0.0030}^{+0.0028}$ & 0.07673$_{-0.00077}^{+0.00075}$ & 0.06$_{-0.14}^{+0.15}$ & 124 & 8.6$_{-1.3}^{+1.6}$& 2.30$_{-0.15}^{+0.16}$ & \\
\hline
TOI~1670~b & 10.98462$_{-0.00051}^{+0.00046}$ & 0.103$\pm$0.002 & 0.59$_{-0.26}^{+0.17}$ & 163.6$_{-53.7}^{+41.7}$ & $<$ 41.3\footnote{\cite{Tran22} give the mass of TOI~1670~b as $<$ 0.13 M$_{J}$, and we have converted this to units of M$_\oplus$ for consistency in this table.} & 2.06$_{-0.15}^{+0.19}$ & \cite{Tran22} \\
TOI~1670~c & 40.74976$_{-0.00021}^{+0.000022}$ & 0.249$\pm$0.005 & 0.09$_{-0.04}^{+0.05}$ & 105.5$_{-29.4}^{+28.6}$ & 200$_{-25}^{+29}$\footnote{\cite{Tran22} give the mass of TOI~1670~c as 0.63$_{-0.08}^{+0.09}$ M$_{J}$. We have converted this into units of M$_\oplus$, for consistency in this table.} & 11.1$\pm$0.28\footnote{\cite{Tran22} give the radius of TOI~1670~c as 0.987$\pm$0.025 r$_{J}$. We have converted this into units of r$_\oplus$, for consistency in this table.} & \\
\hline
TOI~421~b & 5.19672$\pm$0.00049 & 0.0560$\pm$0.0018 & 0.163$_{-0.071}^{+0.082}$ & 128.9$_{-27.2}^{+24.9}$ & 7.17$\pm$0.66 & 2.68$_{-0.18}^{+0.19}$ & \cite{Carl20} \\
TOI~421~c & 16.06819$\pm$0.00035 & 0.1189$\pm$0.0039 & 0.152$\pm$0.042 & 114.7$_{-13.3}^{+15.6}$ & 16.42$_{-1.04}^{+1.06}$ & 5.09$_{-0.15}^{+0.16}$ & \\
\hline
\end{tabular}
\renewcommand{\arraystretch}{1.0}
\end{table}
\begin{table}[bht]
\renewcommand{\arraystretch}{1.5}
\centering
\caption{The key parameters for the host stars of the planetary systems studied in this work. We note that the only stellar parameter used in our simulations is the mass, but we include the other parameters here for completeness. \label{table:host_stars}}
\begin{tabular}{lcccccl}
\hline \hline
Star & Mass (M$_\odot$) & Radius (R$_\odot$) & Spectral Class & Effective Temperature (K) & Age (Gyr) & Reference \\
\hline
HD~15337 & 0.90~$\pm$~0.03 & 0.856~$\pm$~0.017 & K1V & 5125~$\pm$~50 & 5.1~$\pm$~0.8 & \cite{Gand19} \\
HD~21749 & 0.73~$\pm$~0.07 & 0.695~$\pm$~0.030 & K4.5V & 4640~$\pm$~100 & 3.8~$\pm$~3.7 & \cite{Drag19} \\
HD~63433 & 0.98~$\pm$~0.03 & 0.912~$\pm$~0.034 & G2V & 5640~$\pm$~74 & 0.414~$\pm$~0.023 & \cite{Mann20} \\
HD~73583 & 0.73~$\pm$~0.02 & 0.65~$\pm$~0.02 & K4V & 4511~$\pm$~110 & 0.48~$\pm$~0.19 & \cite{Barr22} \\
LTT~3780 & 0.401~$\pm$~0.012 & 0.374~$\pm$~0.011 & M4V & 3331~$\pm$~157 & 3.10$_{-0.98}^{+6.20}$ \footnote{No age is given for LTT~3780 in \cite{Clout20}; the age presented here is taken from the recent characterisation of the star by \cite{Bonf24}.} & \cite{Clout20}  \\
TOI~1670 & 1.21~$\pm$~0.02 & 1.316~$\pm$~0.019 & F7V & 6170~$\pm$~61 & 2.53~$\pm$~0.43 & \cite{Tran22} \\
TOI~421  & 0.852$_{-0.021}^{+0.025}$ & 0.871~$\pm$~0.012 & G9V & 5325$_{-58}^{+78}$ & 9.4$_{-3.1}^{+2.4}$ & \cite{Carl20} \\
\hline
\hline
\end{tabular}
\renewcommand{\arraystretch}{1.0}
\end{table}
\clearpage

\section{Methodology} \label{sec:methods}

To investigate whether the systems herein are tightly dynamically packed, we performed extensive $n$-body simulations using the Hybrid integrator within Mercury \citep{Chambers99}. We build upon our earlier work studying the dynamical stability of known exoplanetary systems \citep[e.g.][]{HornerHUAqr,HornerQSVir,Horner181433,Witt24Sex,Witt73526}, examining a sample space between the two planets in each system that is a regular grid in semi-major axis ($a$) - eccentricity ($e$) - longitude of periastron ($\omega$) - mean anomaly ($M$) space. 

For each system studied (except HD~63433), we carried out a total of 343170 individual simulations, distributed across a regular grid in $a$-$e$-$\omega$-$M$ space, each of which tested the stability of a different `potential Earth' located between the orbits of the two planets known in that system. \change{Since a key focus of current exoplanetary science is the search for Earth-sized planets, we chose to examine scenarios where the injected planet had mass identical to the Earth (i.e. 1 $M_\oplus$). For the systems considered, previous observational studies lack the precision to detect such planets -- and thus it is fair to consider that the absence of evidence for such planets therein is not, necessarily, evidence of absence.} 

For each simulation, the two known planets were allocated their canonical mass and orbital elements (as detailed in Section~\ref{sec:systems}, with TOI~1670~b set to the maximum allowed mass of 41.3 M$_\oplus$, and HD~21749~b set to the estimated mass of 2.5 M$_\oplus$)\change{\footnote{The initial mean anomalies of the two planets in each system were set based on the stated time of transit midpoint (Time of Conjunction) for the system as detailed on the NASA Exoplanet Archive (accessed 19/10/23) for the default parameter set. In each case, we set the mean anomaly of the inner planet in the system to zero, then calculated the mean anomaly for the outer planet based on that planet's orbital period and the time between its transit midpoint and that of the inner planet.}}. \change{In the case of HD~21749~c, no values were available for $e$ and $\omega$, and so these were set to zero in our simulations.}
A hypothetical third planet was then placed on an orbit between the two known planets in the system. That hypothetical world was allocated a mass equal to that of the Earth (i.e. 5.97219 $\times$ $10^{24}$ kg), with an initial orbit placed on a single point in the $a$-$e$-$\omega$-$M$ space studied. 

In total, for each system tested, we sampled 246 unique values of orbital semi-major axis, evenly distributed between the semi-major axes of the inner and outer planets. At each value of semi-major axis tested, we sampled 31 unique eccentricities, between 0 (i.e. a circular orbit) and 0.9 (highly eccentric). For each $a$-$e$ pair, we tested nine unique values of $\omega$, evenly distributed between 0 and 360$^\circ$, and at each of those $a$-$e$-$\omega$ locations, we tested five unique values of $M$, again evenly distributed. This gave a total of 343170 unique test scenarios.

In the case of HD~63433, whilst orbital solutions\change{\footnote{In this work, we use the default parameter set for the HD~63433 system, as detailed by the NASA Exoplanet Archive. This corresponds to the final solutions presented by \cite{Mann20}, in which they fixed these values to zero during their analysis.}} and radii were available for the two known planets, no mass was available for either world. We therefore created three hypothetical versions of the HD~63433 system, using the mass-radius relationship detailed in \citet{ChenKipping}, and illustrated in Figure 3 of that work, to estimate masses for the two planets based on their published radii. In the first hypothetical system, we considered the maximum mass each planet might reasonably be expected to have, based on its published radius. This yielded masses of 18 $M_\oplus$ and 23 $M_\oplus$ for planets HD~63433~b and c, respectively. In the second system, we considered the most likely/nominal mass that would be expected for their radii - yielding 5 $M_\oplus$ and 7 $M_\oplus$. Finally, in the third setup, we considered the lowest reasonable masses - 2 $M_\oplus$ and 2.2 $M_\oplus$. We then carried out an identical suite of simulations to those described above for each of these three versions of the HD~63433 system. 

All simulations were run for a period of $10^6$ years, using a very short time-step (less than 1/40th the orbital period of the innermost planet in the simulation), and with a version of Mercury that has been modified to include first and second order post-Newtonian corrections \citep[as detailed in][]{HornerMilankovic}. If, during the one million year simulation, one of the three planets (two known, one hypothetical Earth) was ejected from the system, collided with another planet, or fell into the central star, the simulation was stopped, and the time of the cataclysmic event was recorded. This allowed a map of the stability of orbits between the two known planets in each system to be created, in a manner similar to our earlier work on the topic \citep[e.g.][]{HornerNNSer,Horner181433,WittAlone}.

\section{Results and Discussion} \label{sec:results}

The bulk results of our suites of simulations are summarised in Table~\ref{table:stable_percentage}, which shows the number of simulations that did not feature a planetary ejection or collision (i.e. proved to be dynamically stable) during the full one million years tested. It is immediately apparent that our sample of \textit{TESS} systems yield dramatically different results in terms of the potential stability of orbits between the two known planets. 

We discuss each system individually below - but note here that our simulations were always likely to include a large number of unstable solutions. We made the conscious decision to test up to high eccentricities, which in turn leads to a large fraction of the scenarios we investigated featuring planets on mutually-crossing orbits. This is a natural recipe for instability, but such scenarios can prove stable if the two planets in question are protected from close encounters through mutual mean-motion resonance\footnote{The poster child for such a scenario is found in the Solar system, with the long-term evolution of Neptune and the dwarf planet (134340) Pluto \citep[e.g.][]{Malhotra95,Malhotra22}. Indeed, examples abound in the Solar system of objects trapped in mean-motion resonance with one or other of the giant planets, whilst moving on orbits that cross those of their host. For more information, we direct the interested reader to the review by \cite{HornerSSRev}, and references therein.}. In general, however, we would expect stable solutions only for simulations with moderate or low eccentricities - an expectation borne out by the results of our simulations. 

\change{In addition, we note that our choice to use test particles of 1 $M_\oplus$ should, in general, have only a relatively small impact on our final results. The gravitational reach of a planet is typically considered in the context of the Hill sphere -- the radius of which is proportional to the cube root of the mass of the planet in question. Increasing the mass of a planet will result in that planet carving out a wider zone of instability around its orbit -- but that effect occurs relatively slowly with mass (as can be seen in our discussion of the results for the HD~63433 system, in Section ~\ref{stability_HD63433}). For this reason, it is fair to consider our results robust across the range of masses that would be allowed by current observational data, since the non-detection of planets between the orbits of those considered in this work already rules out planets that are sufficiently massive as to yield markedly different dynamical results.}

\begin{table}[b]
\renewcommand{\arraystretch}{1.5}
\centering
\caption{The total number of stable simulations featuring an added one M$_\oplus$ planet located between the two known planets in the systems studied in this work (centre column). In total, for each system, 343170 individual trials were attempted. The right hand column shows the percentage of the total number of simulations that proved dynamically stable for each system. For HD~63433, three suites of simulations were attempted, with varying masses for the two known planets therein, as described in detail in Section~\ref{sec:methods}. \label{table:stable_percentage}}
\begin{tabular}{lcc}
\hline \hline
System & Number of Stable Simulations & Percentage of Stable Simulations \\
\hline
HD~15337 & 39811 & 11.6~\% \\
HD~21749 & 21760 & 6.34~\% \\
HD~73583 & 14645 & 4.27~\%\\
LTT~3780 & 109010 & 31.8~\%\\
TOI~1670 & 1 & 2.91$\times$10$^{-4}~$\% \\
TOI~421 & 3808 & 1.11~\% \\
\hline
HD~63433 (High Mass) & 35849 & 10.4~\%  \\
HD~63433 (Medium Mass) & 48025 & 14.0~\% \\
HD~63433 (Low Mass) & 54556 & 15.9~\%\\
\hline
\hline
\end{tabular}
\renewcommand{\arraystretch}{1.0}
\end{table}

\clearpage
\subsection{HD~15337}

\begin{figure*}
\centering
\begin{tabular}{cccccc}
\multicolumn{3}{c}{\includegraphics[width=0.5\textwidth,height=0.3\textheight]{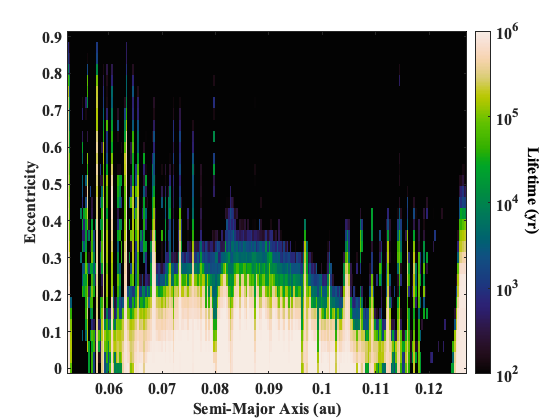}} &\multicolumn{3}{c}{\includegraphics[width=0.5\textwidth,height=0.3\textheight]{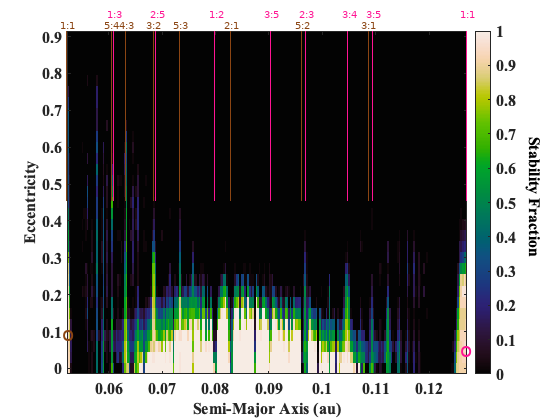}} \\
\end{tabular}
\caption{The dynamical stability of the Earth-mass test particles distributed between the orbits of HD~15337~b and HD~15337~c, as a function of semi-major axis, $a$, and eccentricity, $e$, over the million years of our simulations. Each point in the plot shows the mean lifetime of 45 test particles that began the simulations at that particular ($a,e$) location, which were distributed across a variety of arguments of periastron, $\omega$, and mean anomalies, $M$. The left-hand panel shows the mean lifetime across the region tested, whilst the right shows the fraction of the simulations that began at a given $a$-$e$ ordinate that survived for the full simulation duration. In the right hand panel, the semi-major axis and eccentricity of the two known planets are marked by hollow circles, with the locations of key mean-motion resonances with those planets shown in the top half of the plot. The resonances with the inner planet (HD~15337~b) are shown in brown, whilst those with the outer planet (HD~15337~c) are shown in pink.}
\label{stability_HD15337}
\end{figure*}

The results of our simulations of the stability of an added Earth-mass planet in the HD~15337 system can be seen in Figure~\ref{stability_HD15337}. The left-hand panel of that plot shows the mean lifetime of the injected planet as a function of the semi-major axis and eccentricity of its orbit, with each point being the mean of 45 individual simulations testing a variety of values for the planet's $\omega$ and $M$. 

The system displays a broad region of stability for low and moderate orbital eccentricities, but the influence of both secular and mean-motion resonances can clearly be seen eating into that region of stability. Two distinct areas of stability can be seen on orbits trapped in 1:1 mean-motion resonance with both the inner and outer planets in the system. Such scenarios are cases where the injected planet is safely trapped in mean-motion resonance with the planet at the same location - making it a Trojan or co-orbital companion to the more massive known planet. 

The right-hand plot shows the fraction of simulations at each unique $a$-$e$ pair that remained stable for the full million years of our simulations. Once again, the Trojan/co-orbital solutions can be clearly seen to the left and right hand sides of the plot. The fine resonant structures at the inner edge of the gap between HD~15337~b and c are still visible - narrow bands where a subset of trials remained stable. These resonant scenarios protect the added Earth-mass planet from ejection or collision with HD~15337~b, despite the fact that the added planet is moving on an orbit that crosses that of the known world. Whilst such scenarios are dynamically feasible, it seems unlikely to us that any planet would move on such an orbit in this system, as the high eccentricity would lead to tidal circularisation of the orbit around periastron, dragging both the eccentricity and semi-major axis to smaller values until the resonant protection was broken, and a collision or ejection event would occur.

In total, some 11.6\% of all simulations proved to be stable for HD~15337 -- with the vast majority located at low eccentricity between semi-major axes of ${\sim}0.065$ and ${\sim}0.11$. Aside from narrow bands of instability driven by mean-motion resonance with one or other of the known planets in the system, this whole region exhibits stability on a scale that would clearly allow for the addition of an Earth-mass planet. In other words, the HD~15337 system is \textbf{not} dynamically packed.

\clearpage

\subsection{HD~21749}
\begin{figure*}
\centering
\begin{tabular}{cccccc}
\multicolumn{3}{c}{\includegraphics[width=0.5\textwidth,height=0.3\textheight]{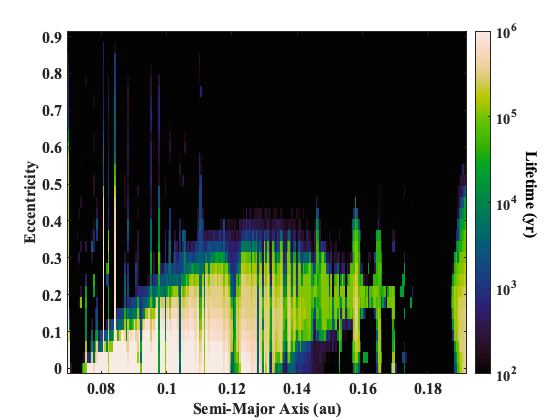}} &\multicolumn{3}{c}{\includegraphics[width=0.5\textwidth,height=0.3\textheight]{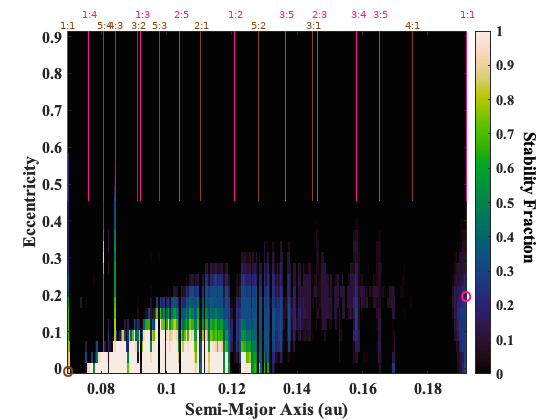}} \\
\end{tabular}
\caption{The dynamical stability of the Earth-mass test particles distributed between the orbits of HD~21749~b and HD~21749~c, as a function of semi-major axis, $a$, and eccentricity, $e$, over the million years of our simulations. Each point in the plot shows the mean lifetime of 45 test particles that began the simulations at that particular ($a,e$) location, which were distributed across a variety of arguments of periastron, $\omega$, and mean anomalies, $M$. The left-hand panel shows the mean lifetime across the region tested, whilst the right shows the fraction of stable simulations in the same area. The right hand panel also shows the location of the two known planets in the system (denoted by hollow circles), and the location of key resonances with those planets, in the top half of the plot.}
\label{stability_HD21749}
\end{figure*}

As detailed in Table~\ref{table:stable_percentage}, some 6.34\% of the simulations for the HD~21749 system proved to be dynamically stable. The distribution of these stable outcomes can be seen in Figure~\ref{stability_HD21749}, with the mean lifetimes from our simulations shown in the left hand panel, and the survival fraction as a function of semi-major axis and eccentricity shown in the right. 

As was the case for HD~15337, the HD~21749 system features a broad area of stability between the orbits of the two known planets, albeit with the great majority of the stable solutions skewed towards smaller semi-major axes and low eccentricities. The simulations again featured a significant number of stable Trojan solutions with the two known planets, but the bulk of the stable solutions were found at semi-major axes between $\sim$0.75 and 0.125 au, at low orbital eccentricities. The more massive planet in the system, HD~21749~c (at a simulated mass of 22.7 M$_\oplus$) has a strong effect clearing the space exterior to 0.125 au of stable solutions.

Interestingly, at semi-major axes greater than 0.125 au, there is a marked tail of stable solutions at moderate eccentricities, with broader spikes of stability at specific semi-major axes (at the location of near-overlapping mean motion resonances with both HD~21749~b and HD~21749~c). This structure is strongly reminiscent of stable populations of objects within the Solar system's asteroid belt, where secular resonances ensure enhanced stability in a region that would otherwise be rendered unstable by the influence of one of the planets (typically Jupiter).

As was the case for HD\,15337, it is clear that the HD\,21749 system is \textbf{not} dynamically packed.  \change{We note that the use of different mass-radius relations to determine the mass of HD\,21749 c would result in a lower estimated mass, and therefore a weaker gravitational influence. Using the relation laid out in \citet{otegi20}, for example, would yield a mass of 0.8-1.5\Mearth, half the value used in our dynamical simulations. Whilst it might seem that this would have a significant impact on the results of our simulations, the results for HD\,63433 show that, in the main, the region of stability is relatively insensitive to changes in the mass of the planets. This is a direct result of the fact that the dynamical reach of a given planet (the Hill sphere) is proportional to the cube root of its mass, coupled with the fact that the two planets themselves are sufficiently widely separated that they are not strongly interacting.}
\clearpage

\subsection{HD~63433}
Since \cite{Mann20} were unable to determine masses for the two planets in the HD~63433 system, we performed three suites of simulations for the system. From one suite to the next, the orbital parameters for HD~63433~b and c were the same -- the sole change was the mass used for the two planets. We ran simulations for a high mass scenario \citep[where both planets had the maximum mass that might be expected for a planet of that diameter, based on the mass-radius relationship described in][]{ChenKipping}; a medium mass scenario (both planets having the most likely/nominal mass that relationship would suggest for their radii); and a low mass scenario (with the lowest masses that the mass-radius relationship would suggest for the planetary radii). 

\begin{figure*}
\centering
\begin{tabular}{cccccc}
\multicolumn{3}{c}{\includegraphics[width=0.5\textwidth,height=0.3\textheight]{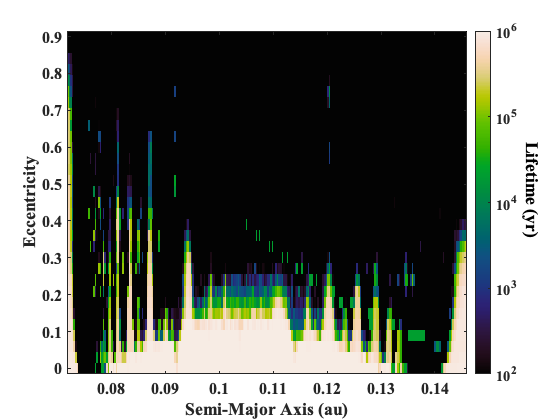}} &\multicolumn{3}{c}{\includegraphics[width=0.5\textwidth,height=0.3\textheight]{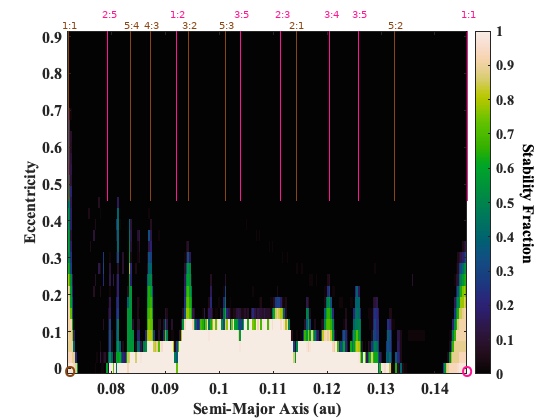}} \\
\multicolumn{3}{c}{\includegraphics[width=0.5\textwidth,height=0.3\textheight]{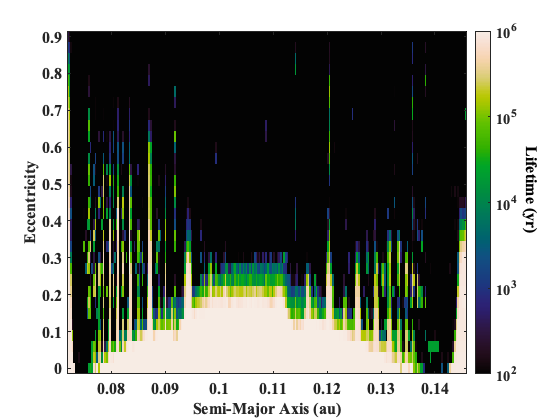}} &\multicolumn{3}{c}{\includegraphics[width=0.5\textwidth,height=0.3\textheight]{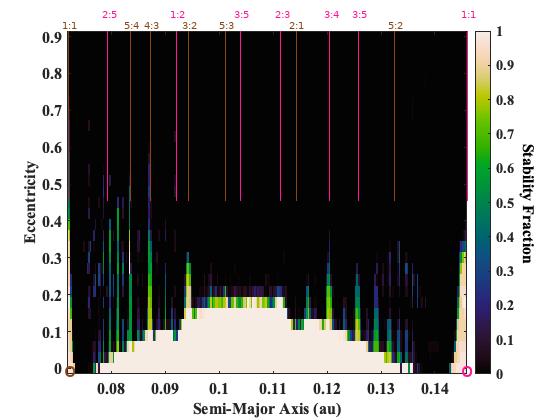}} \\
\multicolumn{3}{c}{\includegraphics[width=0.5\textwidth,height=0.3\textheight]{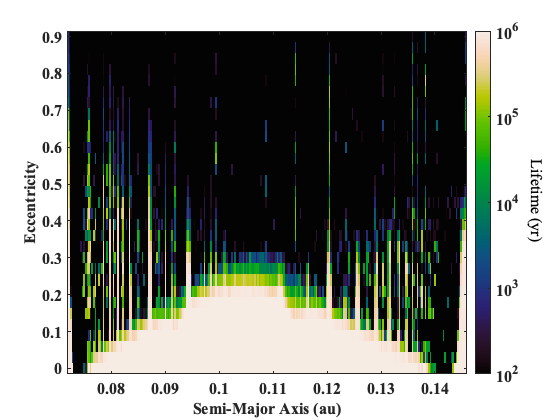}} &\multicolumn{3}{c}{\includegraphics[width=0.5\textwidth,height=0.3\textheight]{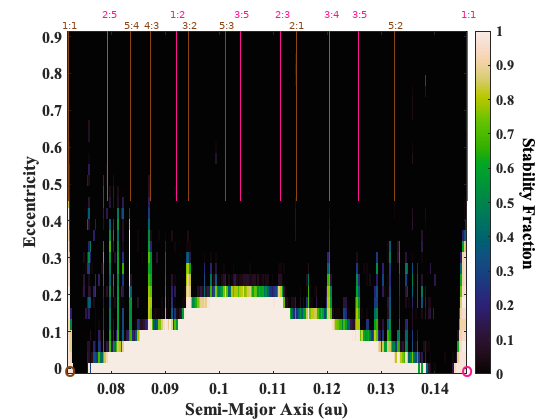}} \\
\end{tabular}
\caption{The dynamical stability of the Earth-mass test particles distributed in the HD~63433 system, as a function of semi-major axis, $a$, and eccentricity, $e$, over the million years of our simulations. Each point in the plot shows the mean lifetime of 45 test particles that began the simulations at that particular ($a,e$) location, which were distributed across a variety of arguments of periastron, $\omega$, and mean anomalies, $M$. The left-hand panels show the mean lifetime across the region tested, whilst the right shows the fraction of stable simulations in the same area. The upper row shows the results for the highest considered masses of HD63433's planets, the middle row shows the intermediate mass solution, and the lower row shows the low mass solution. The locations of the two known planets in the system are marked by hollow circles in the right hand plot, along with the position of key mean-motion resonances with those planets.}
\label{stability_HD63433}
\end{figure*}

All three scenarios tested revealed a broad island of stability between the orbits of HD~63433~b and c. The fraction of stable simulations was highest for the low-mass scenario, and lowest for the high-mass scenario (15.9\% for the low-mass case; 14.0\% for the medium-mass case; and 10.4\% for the high mass scenario). The mean lifetime and stability of our simulations across these three scenarios can be seen in Figure~\ref{stability_HD63433}. In that Figure, the upper panel shows our results for the high-mass scenario, the middle panel is the medium-mass scenario, and the lower panel shows the results for the low-mass scenario.

In all three scenarios, there is a broad island of stability spanning most of the semi-major axis space between the orbits of the two planets, at low and moderate eccentricities. As the mass of the two known planets in the system is increased, the inner and outer edges of that island are pushed slightly away from the orbits of those planets -- an effect that is easier to see at the outer edge of our plots (with the stable island ending just outside 0.13 au for the high-mass case, but at almost 0.14 au for the low mass case. This is a direct result of the `gravitational reach' (i.e. the Hill sphere) of the two known planets increasing as their mass goes up. The effect remains relatively subtle, however, since the scale of the Hill sphere is roughly proportional to the cube-root of the planet's mass - so an increase in mass of a factor of ten (for HD~63433~c) equates to an increase in the radius of that planet's Hill sphere of $\sim$2.15 times. 

In each scenario tested, there are once again a large number of stable scenarios where the added Earth-mass planet moves as a Trojan companion of one of the two planets in the system. The protective influence of mean-motion resonances can be seen in each case in the form of the vertical spikes of stability protruding from the stable island (which are most clearly visible in the fractional stability plots, shown as the right-hand column in Figure~\ref{stability_HD63433}). 

The region where all tested solutions at a given $a$-$e$ location proved dynamically stable (i.e. the area with a stability fraction of 100\%) extends to slightly larger eccentricities for lower masses of the two known planets. This offers an interesting insight into how, in some scenarios, the detection of a new planet and determination of its orbit can be used to constrain the mass of the other planets already known in that system.  \change{This result suggests that, in certain specialised cases, the detection of a new planet and determination of its orbit could be used to constrain the mass of the other planets already known in that system. This is, perhaps, not unprecedented, as previous work has shown more general cases where the discovery of a second planet in a system dramatically alters our knowledge and characterisation of the first planet found \citep[e.g.][]{witt12, trifonov17, nagel19}. }

In this case, should a third planet be found in the HD~63433 system, moving on a moderately eccentric orbit between those of HD~63433~b and c, simply knowing that that planet exists on that orbit would impose a maximum possible mass to the two known planets. Imagine, for example, that the planet is located at a $= 0.1$ au and e $= 0.2$. Such a scenario would suggest that the two known planets must have masses close to the minimum allowed by the \cite{ChenKipping} mass-radius relation, as such an orbit would be unstable for higher mass scenarios.

It is interesting to compare our results with the dynamical tests described in the discovery paper for the HD~63433 system \citep{Mann20}. In that work, the authors found an island of stability between 0.099 and 0.112~au -- significantly narrower than that seen in Figure~\ref{stability_HD63433}. In their simulations, \cite{Mann20} used masses of 5.5 and 7.3 M$_\oplus$ -- essentially identical to the masses used in our simulations. It is unclear from \cite{Mann20} what orbital eccentricities and rotation angles were used in their simulations, but it is clear that their island of stability matches up well with the central block of the stability island seen in Figure~\ref{stability_HD63433}, bracketed by the enhanced instability caused by the 2:1 mean-motion resonance with HD~63433~b (exterior to 0.11 au) and the 1:2 mean-motion resonance with HD~63433~c (just outside 0.09 au). That region also features two spikes of enhanced stability caused by the 3:2 resonance with HD~63433~b at $\sim$0.095 au and the 2:3 resonance with HD~63433~c at $\sim$0.11 au.

It is clear from our results, and those detailed in \cite{Mann20}, that, as was the case for the two previous systems discussed in this work, the two planets in the HD~63433 system are \textbf{not} tightly dynamically packed, and that there is ample room for a third, as yet undetected, planet to orbit between them. 
\clearpage
\subsection{HD~73583}
\begin{figure*}
\centering
\begin{tabular}{cccccc}
\multicolumn{3}{c}{\includegraphics[width=0.5\textwidth,height=0.3\textheight]{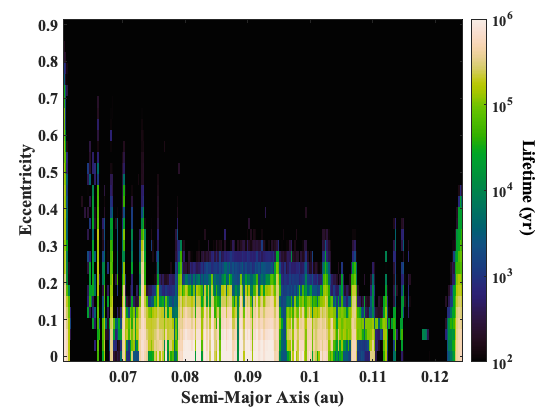}} &\multicolumn{3}{c}{\includegraphics[width=0.5\textwidth,height=0.3\textheight]{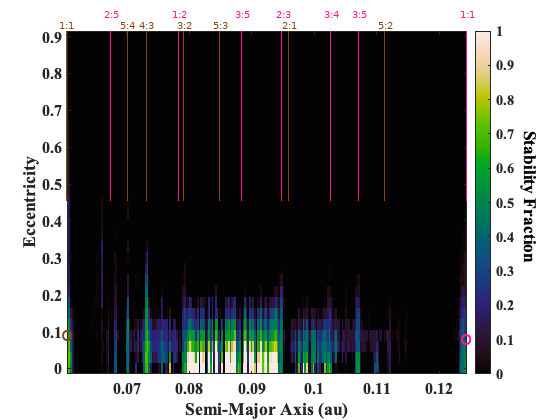}} \\
\end{tabular}
\caption{The dynamical stability of the Earth-mass test particles distributed in the HD~73583 system, as a function of semi-major axis, $a$, and eccentricity, $e$, over the million years of our simulations. Each point in the plot shows the mean lifetime of 45 test particles that began the simulations at that particular ($a,e$) location, which were distributed across a variety of arguments of periastron, $\omega$, and mean anomalies, $M$. The left-hand panel shows the mean lifetime across the region tested, whilst the right shows the fraction of stable simulations in the same area, along with the location of the two known planets, and several of their mean-motion resonances.}
\label{stability_HD73583}
\end{figure*}

The results of our simulations of the HD~73583 system can be seen in Figure~\ref{stability_HD73583}. In total, just 4.27\% of all simulations proved stable for the full one million years of integration time. Despite this relatively low number, we can once again see both evidence of stable Trojan solutions for Earth-mass planets trapped in 1:1 mean-motion resonance with both HD~73583~b and c, as well as a broad island of stability between the orbits of the two planets.

In this case, the core region of the stable island is significantly smaller and more compact than those seen for the previous systems studied. It has a sharp inner edge just interior to 0.08 au and a sharp outer edge at around 0.095 au - locations that correspond to the position of the 1:2 mean motion resonance with HD~73583~c (inner edge) and the 2:1 resonance with HD~73583~b (outer edge). Even in this core region, there are narrow strips of instability tied to higher-order resonant interactions with both HD~73583~b and c. This main island is accompanied by two smaller regions where some, but not all, of the simulations proved to be stable -- including distinct spikes of resonance-induced stability (such as at the locations of the 4:3 and 3:5 mean-motion resonances with the two known planets). \change{It is worth noting that \cite{EM23} recently reexamined this system, finding no evidence for additional planets. In that work, they point out that the star exhibits significant levels of stellar activity, with a period that would be similar to the orbital period of any planet within the stable island described in this work. Such activity may well render the search for such a planet particularly challenging.} 

Of all systems tested in this work, HD~73583 is the one where our dynamical studies place the tightest constraint on the range of orbits a stable planet could occupy between the two known exoplanets. Once again, we find the this system is quite clearly \textbf{not} dynamically packed. 

\clearpage
\subsection{LTT~3780}
The LTT~3780 system proved to have the greatest number of stable outcomes of those we study in this work, with $\sim$31.8\% of all simulations remaining stable for the full million years of our integrations (as can be seen in Table~\ref{table:stable_percentage}. This is not a surprise, however, as this is the system that features the two planets that are most widely spaced. LTT~3780~b moves on an orbit with a period of ${\sim}0.768$ days, whilst LTT~3780~c has an orbital period of ${\sim}12.3$ days -- a factor of 16 times longer. If one were to imagine a tightly packed system of planets all trapped in mutual 1:2 mean-motion resonance with their neighbours (an extension of the Laplace resonance seen with the Jovian moons Io, Europa and Ganymede, whose periods are locked in 1:2:4 resonance), one could imagine a chain of 1:2:4:8:16 -- in other words, a scenario where three additional planets could be found between those observed in this system! Whilst this might seem an extremely unlikely scenario, we note that several such systems have already been discovered. The Gliese 876 system features just such a Laplace resonance \citep[with planets e, b, and c, as described in][]{GJLaplace}, with more extreme examples including TOI~178 \citep[with five mutually resonant planets, and a sixth in near-resonance with the chain][]{Leleu21} and HD~110067 \citep[which features a resonant chain of six planets][]{Luque23}.

\begin{figure*}
\centering
\begin{tabular}{cccccc}
\multicolumn{3}{c}{\includegraphics[width=0.5\textwidth,height=0.3\textheight]{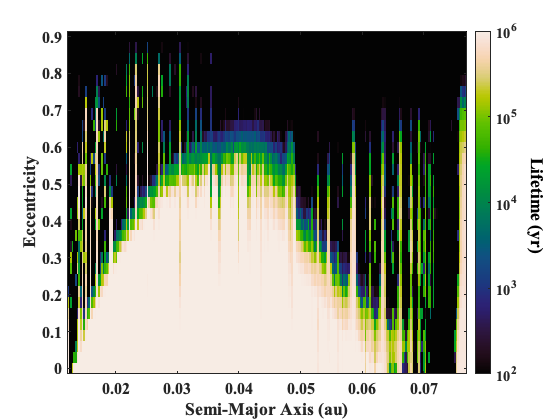}} &\multicolumn{3}{c}{\includegraphics[width=0.5\textwidth,height=0.3\textheight]{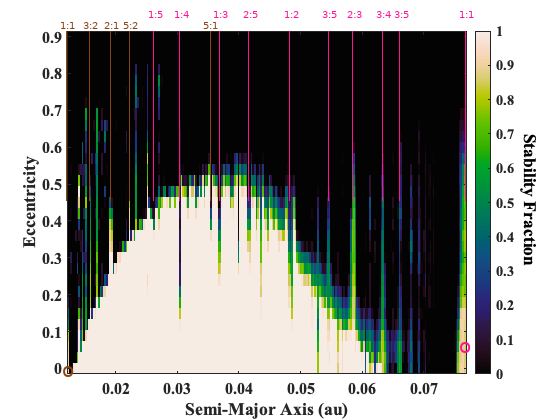}} \\
\end{tabular}
\caption{The dynamical stability of the Earth-mass test particles distributed in the LTT~3780 system, as a function of semi-major axis, $a$, and eccentricity, $e$, over the million years of our simulations. Each point in the plot shows the mean lifetime of 45 test particles that began the simulations at that particular ($a,e$) location, which were distributed across a variety of arguments of periastron, $\omega$, and mean anomalies, $M$. The left-hand panel shows the mean lifetime across the region tested, whilst the right shows the fraction of stable simulations in the same area, along with the location of the two known planets and their mean-motion resonances.}
\label{stability_LTT7380}
\end{figure*}

In our simulations, of course, we only tested the presence of a single additional planet in each run. The result was a vast area of phase space where that planet could move on a stable, dynamically feasible orbit. The results shown in Figure~\ref{stability_LTT7380} bear this out. In addition to the potential presence of Trojan companions, the stable region stretches from just outside the orbit of LTT~3780~b to approximately 0.063au, with additional small regions of stability beyond that tied to mean motion resonances. 

Those regions of potential stability enabled by mean-motion resonances are particularly impressive for this system, with stable solutions stretching up through almost the full range of orbital eccentricities tested in this work (with stable solutions at e $> 0.8$ in three resonant spikes between 0.02 and 0.03 au). Quite simply, of all the systems studied in this work, LTT~3780 has the most room for additional planets between the two that are already known, and is clearly far from being a dynamically packed system based on our current knowledge.
\clearpage
\subsection{TOI~1670}
\begin{figure*}
\centering
\begin{tabular}{cccccc}
\multicolumn{3}{c}{\includegraphics[width=0.5\textwidth,height=0.3\textheight]{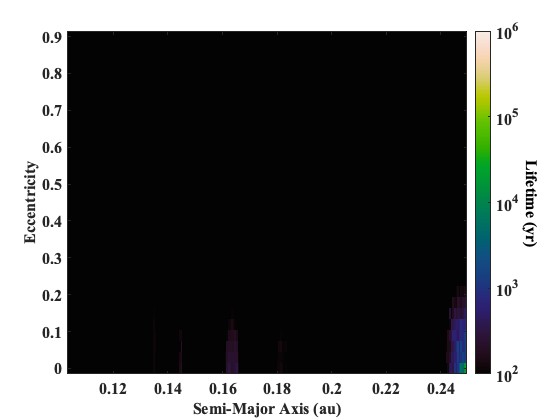}} &\multicolumn{3}{c}{\includegraphics[width=0.5\textwidth,height=0.3\textheight]{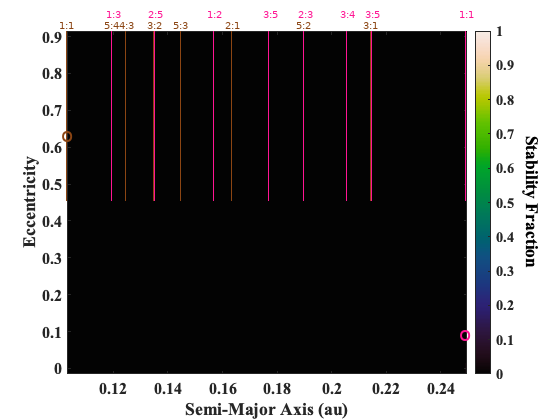}} \\
\end{tabular}
\caption{The dynamical stability of the Earth-mass test particles distributed between the orbits of TOI~1670~b and TOI~1670~c, as a function of semi-major axis, $a$, and eccentricity, $e$, over the million years of our simulations. Each point in the plot shows the mean lifetime of 45 test particles that began the simulations at that particular ($a,e$) location, which were distributed across a variety of arguments of periastron, $\omega$, and mean anomalies, $M$. The left-hand panel shows the mean lifetime across the region tested, whilst the right shows the fraction of stable simulations in the same area. The location of the two known planets, and their mean-motion resonances, are shown in the right hand panel.}
\label{stability_TOI1670}
\end{figure*}
In stark contrast to LTT~3780, the TOI~1670 system proved to be by far the least hospitable to the presence of additional planets. Indeed, of 343170 individual simulations carried out of this system, just one remained stable until the end of the one million year simulation -- a stable fraction of just 2.91$\times$10$^{-4}$\%\footnote{\change{The solution that remained stable for the full simulation featured an Earth-mass planet that had an initial orbit with $a$ = 0.249298 au, $e$ = zero, $\omega$ = 311.9999985$^\circ$, and $M$ = 252$^\circ$. Given the extreme instability of the rest of the ensemble tested for this system, it would be interesting in future to perform lengthier simulations for this single scenario, to assess the degree to which such a planet could exist in the system as it is currently configured, but such simulations are beyond the scope of the current work.}}. This is not a surprise -- as we note in Section~\ref{subsec:1670}, this system features by far the most massive planets considered in this work (which therefore have by far the largest gravitational reach). In addition, the inner of those planets, TOI~1670~b, moves on a highly eccentric orbit, with e = 0.59$_{-0.26}^{+0.17}$. The result of the planet's high orbital eccentricity, coupled with the high mass of the outer planet, TOI~1670~c, is to render the entire area between the two known planets unstable. As a result, we can conclude that the TOI~1670 system \textbf{is} dynamically packed -- in other words, there is simply no room between the two known planets for any others to exist, based on our current knowledge of the planets involved.
\clearpage

\subsection{TOI~421}
Our simulations of the potential stability of an additional planet in the TOI~421 system proved to be highly unstable, with just $\sim$1.11\% of all simulated systems surviving for the full one million years investigated in this work. As with TOI~1670, this level of instability is not a great surprise. The two known planets in the TOI~421 system move on moderately eccentric orbits -- meaning that they approach one another more closely than might be expected from a quick glance at their orbital periods or semi-major axes. 

\begin{figure*}
\centering
\begin{tabular}{cccccc}
\multicolumn{3}{c}{\includegraphics[width=0.5\textwidth,height=0.3\textheight]{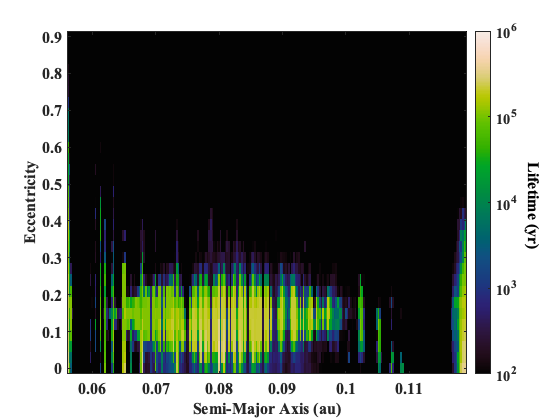}} &\multicolumn{3}{c}{\includegraphics[width=0.5\textwidth,height=0.3\textheight]{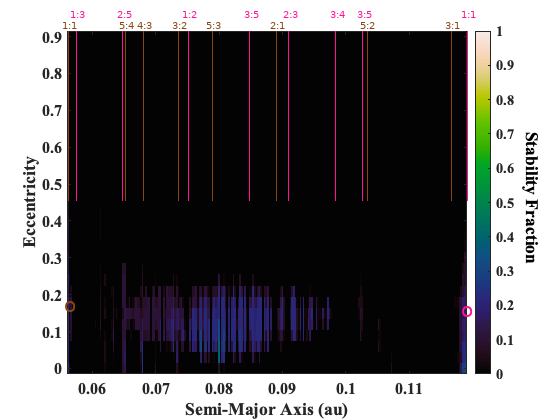}} \\
\end{tabular}
\caption{The dynamical stability of the Earth-mass test particles in the TOI~421 system, as a function of semi-major axis, $a$, and eccentricity, $e$, over the million years of our simulations. Each point in the plot shows the mean lifetime of 45 test particles that began the simulations at that particular ($a,e$) location, which were distributed across a variety of arguments of periastron, $\omega$, and mean anomalies, $M$. The left-hand panel shows the mean lifetime across the region tested, whilst the right shows the fraction of stable simulations in the same area, along with the locations of the two confirmed planets in the system, and their mean-motion resonances.}
\label{stability_TOI421}
\end{figure*}

The stability of Earth-mass test particles between the two planets in the TOI~421 system is plotted in Figure~\ref{stability_TOI421}. Whilst there is a broad region between the two planets where simulations proved stable for mean lifetimes measured in tens or hundreds of thousands of years, very few of those simulations actually survived for the full million year simulation time. The result is clearly seen in the stability fraction plot to the right of Figure~\ref{stability_TOI421} - a relatively broad area, between $\sim$0.65 and $\sim$0.98 au where a small fraction ($<$~30\%) of simulations at a given location proved dynamically stable on those timescales. Almost all of those solutions require moderately eccentric orbits, with 0.05 $<$ e $<$ 0.25. 

Whilst our simulations cannot definitively rule out the presence of an unseen planet between the orbits of TOI~421~b and TOI~421~c, they are strongly suggestive that no such planet exists. As such, we consider that the planets in the TOI~421 system are \textbf{likely} to be tightly dynamically packed. However, if future observations reveal that the orbital eccentricities of those two planets are lower than the best-fit values considered in that work, that would likely increase the potential for stable orbital solutions between the two. As such, it would be interesting to revisit the orbital solutions for the TOI~421 system in the years to come, to see whether the addition of new observations alters the best-fit solutions for the two known planets.

\subsection{Comparison with DYNAMITE} 
\change{It is interesting to compare our results to those obtained by \cite{Dynamite}, who studied three of the systems in this work using their DYNAMITE package. DYNAMITE follows a significantly different approach to identifying the potential for additional unseen planets to exist in a given planetary system to that described in this work. Where we consider purely the dynamical stability of the system, and search for islands of stability based on detailed $n$-body simulations, DYNAMITE considers a simpler criterion for stability -- simply requiring that any two planets be separated by at least eight mutual Hill radii to be considered stable. This allows for a far more rapid assessment of the degree to which a system is ‘dynamically packed’, but has the disadvantage that it could miss potentially stable scenarios that are very tightly packed (with stability ensured by mean-motion resonance; such as the potential ‘exoTrojans’ discussed earlier in this work) and regions where two widely-spaced planets would be rendered unstable due to resonant interactions. On the flip side, our methodology purely searches for regions of stability where planets could exist, whilst DYNAMITE produces predictions with far more detail -- not only showing where planets could exist, but also predicting their most likely orbital period, mass, and radius -- information that is of great use to researchers attempting to find the predicted planets. \cite{Dynamite} examined three of the systems considered in this work -- HD~15337 (as TOI-402), HD~63433 (as TOI-1726) and LTT~3780 (as TOI-432). They predict that each of these systems could well host at least one undetected planet in the gap between the two already known therein.}

\change{For HD~15337, \cite{Dynamite} predict a planet with a radius of 1.88 R$_\oplus$ with an orbital period of 9.04 days (using their period ratio model), or of the same size with an orbital period of 6.21 days (using their clustered periods model). These correspond to semi-major axes of 0.082 and 0.0638 au, respectively. The first of these falls just inside the location of the 2:1 mean motion resonance with HD~15337~b - a region that is stable for orbits with low eccentricity. The second lies just interior to the location of the 4:3 MMR with HD~15337~b -- a region that is dynamically unstable. However, given the width of the 4:3 resonance at this location, it seems plausible that such a planet could, just, be rendered stable by that resonance - but that a period ratio of 4:3 (i.e. a period of $\sim$6.34 days, or semi-major axis of 0.06473 au) would be much more likely to render such a scenario feasible.}

\change{For the HD 63433 system, \cite{Dynamite} suggest an additional planet with a period of either 9.34 days or 12.1 days – corresponding to semi-major axes of 0.08622 and 0.10246 au. Both of these solutions agree well with our dynamical mapping (as seen in Figure 4), so long as the orbital eccentricities of the proposed planets are low (less than ~0.1). In this case, then, our results and those from DYNAMITE are in good agreement.}

\change{For LTT3780, \cite{Dynamite} suggest an additional planet with a period of either 1.06 or 22.6 days -- corresponding to semi-major axes of 0.015005 or 0.11537 au. The innermost of these planets falls within the island of stability shown in Figure 6, and so is in agreement with our findings, so long as its orbital eccentricity was relatively low. The outer of the two proposed solutions is exterior to the orbit of LTT~3780~c, and so lies beyond the region considered in this work. We note, however, that that solution lies slightly interior to the 1:2 mean-motion resonance with that planet - and so it would be worth investigating whether such a solution would be stable against the potentially disruptive influence of that resonance (a feature that can be seen in most of the dynamical maps presented in this work, across several of the systems considered).}

\subsection{Stability on longer timescales?}

\change{It should be noted that the million year duration of our simulations is between two and four orders of magnitude shorter than the ages of the systems being studied. The computational requirements for extended $n$-body simulations are considerable, naturally truncating the time scales over which such simulations can be reasonably carried out. This is a widely recognised challenge for the dynamical investigation of the stability of systems such as these and, as a result, a number of authors have proposed techniques by which a system's stability can be estimated on much longer timescales \citep[e.g.][]{SPOCK, VM20}. Additional diagnostics include the use of chaos indicators \citet{cincotta1999,cincotta2000} that measure the divergence of orbits and have been applied to exoplanetary systems \citep{gozdziewski2001a,gozdziewski2002,satyal2013,satyal2014}.} 

\change{Whilst it would be interesting in the future to compare the results of simulations such as ours to the longer term predictions that such techniques can make, we note that the boundaries between stable and unstable behaviour seen in our simulations are often very sharp, marked by a rapid change in both mean lifetime and stability fraction.  Such behaviour increases the likelihood that the great majority of the area encompassed by the broad islands of stability identified in this work would prove stable on timescales far longer than those studied by our simulations, and we would expect that this would be reflected in the results that would be obtained using these alternative stability analysis techniques.}

\clearpage
\section{Conclusions} \label{sec:conclusion}
In this work, we have carried out an in-depth dynamical study of seven planetary systems discovered by the \textit{Transiting Exoplanet Survey Satellite, TESS}, to attempt to determine which, if any, of those systems contain planets that are sufficiently widely spaced as to permit an additional, unseen, planet between their orbits. This work is motivated by the growing consensus that a significant fraction (or even the great majority) of exoplanetary systems containing multiple transiting planets are dynamically packed -- that is, that the planets therein are sufficiently close together that their dynamical interactions preclude the presence of any addition planets `stuffed into the gaps' between them. 

To do this, we ran 343170 unique simulations for each planetary system considered, each of which featured the addition of an Earth-mass planet moving on an orbit between those of the two planets known in that system. These Earth-mass planets were distributed according to a regular grid in semi-major axis, eccentricity, argument of pericentre and mean anomaly, such that the space between the two known planets in a given system could be thoroughly sampled.
Individual simulations followed the three planets (two real, one hypothetical) for a period of one million years -- with the simulation stopping if any of the planets were ejected, or collided with either another planet or the central star. If such an event happened, the time at which the collision or ejection occurred was recorded, and the simulation was brought to a halt.

In this way, we were able to generate dynamical stability maps for each of the systems tested, allowing us to determine whether it is plausible that an addition, unseen planet moves in those systems, between the two confirmed planets therein. Our results were as follows:

\begin{itemize}
\item The HD~15337 system \citep[with planets discovered by][]{Gand19} shows a broad island of stability between the orbits of the two known planets, and so could readily host an additional planet in that space. In other words, the planets already known in that system are \textbf{not tightly packed}.
\item The HD~21749 system \citep[planets discovered by][]{Drag19} also shows a broad stable region, particularly between $a\sim$0.75 and 0.125~au, truncated by a narrow band of resonance-induced instability at $\sim$0.12 au. In other words, the planets already known in that system are \textbf{not tightly packed}. 
\item The HD~63433 system \citep[planets discovered by][]{Mann20} has a very broad island of stability spanning almost the entire space between the two known planets. In this system, we tested three scenarios in which the masses of the known planets were varied by an order of magnitude, to span the plausible masses afforded by the known radii of HD~63433~b and c. The broad island of stability was present in all scenarios, albeit being slightly larger for the lower-mass simulations. In other words, the planets already known in the HD~63433 system are \textbf{not tightly packed}, and there is ample space for at least one additional planet between their orbits.
\item The HD~73583 system \citep[planets discovered by][]{Barr22} features a relatively small island of stability between ${\sim}0.08$ and 0.095 au, with additional small stable features on either side. There is definitely room in the system for an additional planet between the two that are currently known, but the potential orbits for that planet are more tightly constrained than for the other loosely packed systems studied in this work. For clarity, therefore, we conclude that the planets in the HD~73583 system are \textbf{not tightly packed}.
\item The LTT~3780 system \citep[planets discovered by][]{Clout20} has the most widely spaced planets of all systems considered in this work, with the outer planet, LTT~3780~c, having an orbital period approximately 16 times longer than the inner planet. As a result, our simulations reveal a vast island of stability between the orbits of the two planets, and it seems likely that there is room for multiple additional planets between the orbits of the two that are currently known. In other words, the LTT~3780 system is \textbf{definitely not tightly packed}.
\item The TOI~1670 system \citep[planets discovered by][]{Tran22} contains the most massive planets of all systems studied in this work, with the innermost, TOI~1670~b, moving on a highly eccentric orbit. As a result, it is not surprising that the system shows no island of stability. As a result, we can confidently state that the planets in the TOI~1670 system are \textbf{definitely tightly packed}. 
\item The TOI~421 system \citep[planets discovered by][]{Carl20} contains two planets on moderately eccentric orbits. A small fraction of our simulations ($\sim$1.11\%) proved stable for the full million years of our simulations. However, all locations in $a$-$e$ space that displayed stability were stable in less than $\sim$30\% of cases. As such, whilst our results do not explicitly rule out the presence of an unseen planet between the two known planets in the TOI~421 system, they suggest that the two planets therein are likely \textbf{tightly packed}. 
\end{itemize}

As technology improves, we will gradually be able to probe the planetary systems identified by \textit{TESS} to ever smaller masses and planetary radii. As such, it is useful to identify those systems in which there is room for additional short period planets. Our results identify five such systems. In addition, we look forward to seeing whether future RV observations of the TOI~421 system confirm the moderate eccentricities of the two planets known therein. If such observations reveal those planets to be moving on more circular orbits than those used in this work, then it may prove plausible for an additional unseen planet to lurk between them. Based on the current best fit solutions, however, we consider it highly unlikely that such a planet could exist in the TOI~421 system, and believe that our results can definitely exclude the existence of any such planet in the TOI~1670 system.

\begin{acknowledgments}
\section*{Acknowledgements}
The authors respectfully acknowledge the traditional custodians of all lands throughout Australia, and recognise their continued cultural and spiritual connection to the land, waterways, cosmos, and community. We pay our deepest respects to all Elders, ancestors, and descendants of the Giabal and Jarowair nations, upon whose lands the Toowoomba campus of the University of Southern Queensland is situated. 
\end{acknowledgments}

\vspace{5mm}
\facilities{The `Fawkes' supercomputing cluster at the University of Southern Queensland was used for all dynamical simulations described in this work}


\software{Mercury \citep{Chambers99}}

\bibliography{Chains}{}
\bibliographystyle{aasjournal}



\end{document}